\documentclass[aps,prb,reprint,superscriptaddress]{revtex4-1}
\usepackage{graphicx}

\begin{document}


\title{
Magnetic shape-memory effect in SrRuO$_3$}

\author{S. Kunkem\"oller}
\affiliation{$I\hspace{-.1em}I$.\ Physikalisches Institut,
Universit\"at zu K\"oln, Z\"ulpicher Str. 77, D-50937 K\"oln,
Germany}

\author{D. Br\"uning}
\affiliation{$I\hspace{-.1em}I$.\ Physikalisches Institut,
Universit\"at zu K\"oln, Z\"ulpicher Str. 77, D-50937 K\"oln,
Germany}

\author{A. Stunault}
\affiliation{Institut Laue Langevin, 71 Avenue des Martyrs, CS 20156, F-38042 Grenoble Cedex 9, France}

\author{A. A. Nugroho}
\affiliation{ Faculty of Mathematics and Natural Science, Institut Teknologi Bandung, Jalan Ganesha 10, 40132 Bandung Indonesia}

\author{T. Lorenz}
\affiliation{$I\hspace{-.1em}I$.\ Physikalisches Institut,
Universit\"at zu K\"oln, Z\"ulpicher Str. 77, D-50937 K\"oln,
Germany}

\author{M. Braden}\email[e-mail: ]{braden@ph2.uni-koeln.de}
\affiliation{$I\hspace{-.1em}I$.\ Physikalisches Institut,
Universit\"at zu K\"oln, Z\"ulpicher Str. 77, D-50937 K\"oln,
Germany}





\date{\today}

\begin{abstract}
Like most perovskites, SrRuO$_3$ exhibits structural phase transitions associated with rotations of the RuO$_6$ octahedra. The application of moderate magnetic fields in the ferromagnetically ordered state allows one to fully control these structural distortions, although the ferromagnetic order occurs at six times lower temperature than the structural distortion. Our neutron diffraction and macroscopic measurements unambiguously show that magnetic fields rearrange structural domains, and that for the field along a cubic [110]$_c$ direction a fully detwinned crystal is obtained.
Subsequent heating above the Curie temperature causes a magnetic shape-memory effect, where the initial structural domains recover.
\end{abstract}


\maketitle


The discovery of the type-II multiferroics has initiated strong interest in the control of magnetism by electric fields as well as in its inverse, the control of a structural distortion by magnetic fields~\cite{Khomskii2009}. While commonly the combination of magnetism and ferroelectric order is considered in the field of multiferroics, the control of structural distortions by magnetic fields is not restricted to polar distortions. Here, we show that indeed also an antiferrodistortive structural distortion, the oxygen octahedron rotation occurring in most perovskites, can be driven by weak magnetic fields in SrRuO$_3$ due to a strong spin-orbit coupling. The manipulation of antiferrodistortive structural domains by magnetic fields opens new paths in magnetoelectric or magnetoelastic coupling in oxides, but it has also to be taken into account when analyzing the magnetic field dependence of any physical property in single crystalline or thin-film SrRuO$_3$~\cite{Koster2012}.

Materials showing strong shape changes in response to an applied magnetic field are well known among Heusler alloys, in which the field either induces a structural phase transition or rearranges martensitic domains~\cite{Ullakko1996,Planes2010,Planes2009}.
The fact that the initial shape can be recovered by reducing the magnetic field or by heating is called magnetic shape-memory effect.
In oxides these phenomenona are very rare~\cite{A.N.Lavrov2002}, but we discovered the domain rearrangement as well as the shape-memory effect in SrRuO$_3$ \cite{Randall1959}.
This perovskite exhibits many fascinating physical properties \cite{Koster2012}, such as an invar effect of the thermal expansion \cite{Kiyama1996}, a linear resistivity that breaks the Ioffe-Regel limit at moderate temperature \cite{Allen1996} and an anomalous Hall effect\cite{Izumi1997} that has been  linked to the existence of magnetic monopoles in momentum space \cite{Fang2003} but is still intensely debated \cite{Kats2004,Haham2011,Koster2012,Itoh2016}. Many of the experimental studies of SrRuO$_3$ are performed using thin-film samples, because of the lack of high-quality single crystals \cite{Koster2012}. Only recently such crystals became available by the floating-zone growth technique in a mirror furnace \cite{Kikugawa2015,Kunkemoeller2016}.



 \begin{figure}
  \includegraphics[width=0.9\columnwidth]{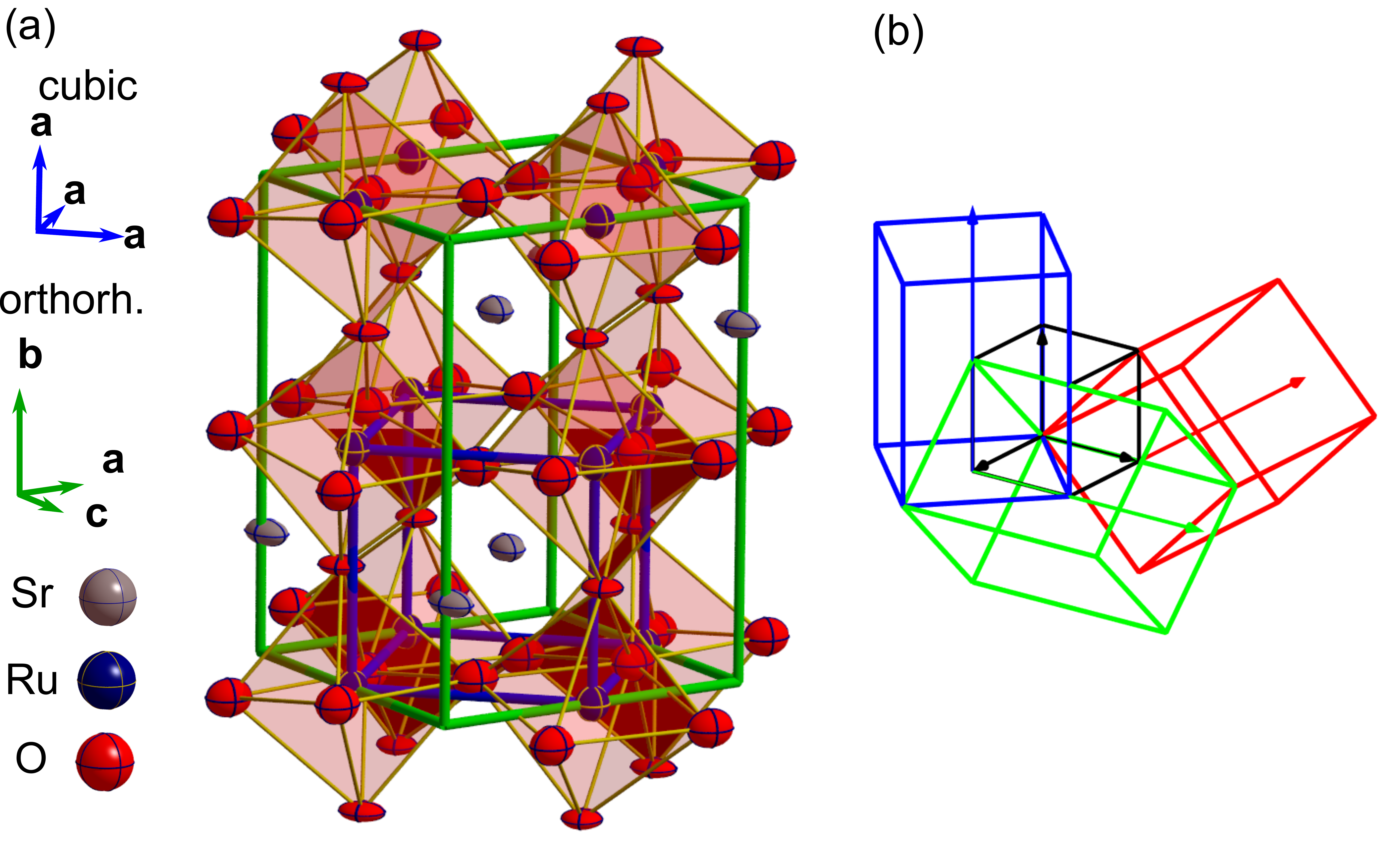}
  \caption{\label{structure}(a) Crystal structure of SrRuO$_3$ at 10~K as obtained from our single-crystal neutron-diffraction analysis.  (b) Illustration of three orthorhombic twin orientations (colored) with their longest axis parallel to one of the cubic axes (black). }
  \label{struc}
 \end{figure}

The high-temperature cubic structure of SrRuO$_3$~\cite{Cuffini1996} transforms upon cooling to a tetragonal phase at 975\,K and at 800\,K  to an orthorhombic phase with space group Pnma \cite{Randall1959,Jones1989,Chakoumakos1998}, see Fig.~\ref{structure} (a). The symmetry reduction from cubic to orthorhombic results in six twin domain orientations that mimic the original cubic symmetry.
The orthorhombic lattice parameters $a$, $b$, $c$ are related to the cubic $a_{c}\simeq 3.93\,$\AA\ via $\vec{a}\|[101]_{c}$, $\vec{b}\|[010]_{c}$, and $\vec{c}\|[\overline{1}01]_{c}$ with $a\sim c \sim \sqrt 2{a}_{c}$ and ${b}\sim 2{a}_{c}$\cite{Pnma}. The four-fold axis of the tetragonal intermediate phase is $b$ and can align along one of the cubic $[100]_c$, $[010]_c$, or $[001]_c$ directions, as is illustrated in Fig.~\ref{struc} (b). For each of these three arrangements, the orthorhombic $a$ and $c$ axes may interchange leading to a total of six twin domain orientations\cite{Pnma}.

\begin{table}
\caption{\label{tab:struct}Twinning fractions at 10 and 170~K. The values in brackets indicate the error on the last digits.}
\begin{tabular}{l  c  c  c  c c c}
\hline \hline
  Twin  & 1             & 2   & 3  & 4 & 5 & 6 \\ \hline
 \textbf{10 K} & 0.286(8) & 0.325(4) & 0.122(4) & 0.134(4) & 0.065(3) & 0.067(3) \\

 \textbf{170 K} & 0.282(7) & 0.329(4) & 0.125(4) & 0.131(4) & 0.066(3) & 0.067(3)
   \\ \hline \hline

\end{tabular}
\end{table}

Large single crystals of SrRuO$_3$ were grown by the floating-zone technique as described in Ref.~\onlinecite{Kunkemoeller2016}, from which we cut samples
of nearly regular cubic shape with the edges aligned along the orthorhombic directions. Hence, four faces of the samples are cubic $(110)_c$ faces and two are cubic $(100)_c$ faces.  The crystal structure of such a cube with edge length of 3.2~mm was analyzed using the neutron diffractometer D9 at ILL \cite{D9}.  The lattice-constant differences of the orthorhombic axes are too small to be resolved by the used instrument, thus contributions of all six twins were superposed. The structural refinement is performed in space group Pnma with Jana 2006 \cite{Petricek} taking into account all six possible twin domain orientations as an incoherent superposition of intensities\cite{StructuralParametersSrRuO3}.  The refinement reveals the crystal to be twinned with highly unequal twinning fractions, see table I. The twinning fractions obtained at both temperatures agree within the error bars underlining the reliability of the structural refinements including six twin domains and the proper treatment of the twin laws \cite{D9,StructuralParametersSrRuO3}.

 \begin{figure}
  \includegraphics[width=0.78\columnwidth]{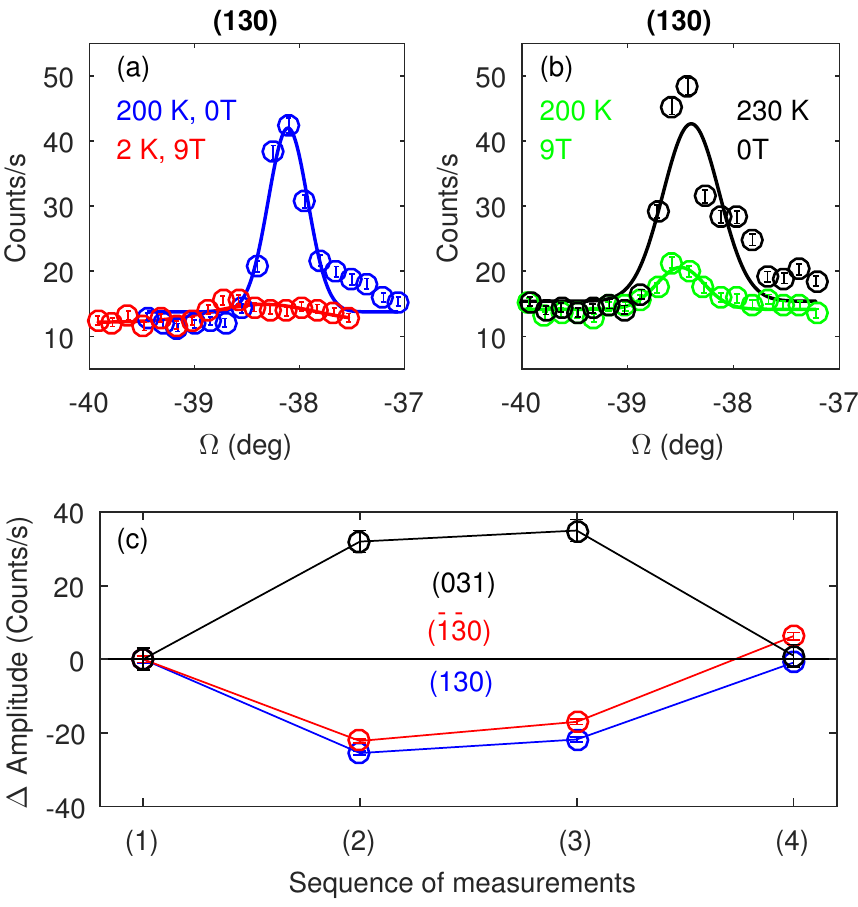}%
  \caption{\label{Twins} Intensities of the $(130)$, $(031)$ and $(\overline{1}\overline{3}0)$ reflections at various conditions. (a) and (b) show rocking scans of the $(130)$ reflection at the starting point of the experiment at 200~K and 0~T (1), after the sample was cooled down to 2~K in 9~T (2), after heating the sample to 200~K maintaining 9~T (3), and in the end after reducing the magnetic field to 0~T and heating to 230~K (4). (c) shows the difference between the amplitude at the starting point of the experiment at 200~K and 0~T (1) and the successive steps (2)-(4). Amplitudes are obtained by fitting gaussians with a constant background.}
 \end{figure}

The twinning fractions and their changes in a magnetic field were investigated at the ILL D3 lifting-counter diffractometer using a crystal with an edge length of 2.5~mm \cite{DOI-D3}. This sample was detwinned beforehand applying the uniaxial-pressure procedure described in Ref.~\onlinecite{Kunkemoeller2016} and was aligned with the magnetic easy axis $c$ parallel to the magnetic field. We analyzed the intensities of several Bragg reflections in order to determine the twinning ratios. The uniaxial-pressure detwinning resulted in a sample with 85 volume~\% of the desired orientation, but the other twins could be clearly detected in our neutron diffraction experiment. In the magnetic field and temperature dependent study we focused on the $(031)$, $(130)$ and $(\overline{1}\overline{3}0)$ reflections because of their sizable intensity and accessibility with the lifting-counter diffractometer. Since the $(130)$ reflection is forbidden in space group Pnma, its intensity results from the $(031)$, $(2\overline{1}\overline{1})$, $(\overline{1}\overline{1}2)$, $(112)$ and $(211)$ reflections of the other 5 twin orientations. These 5 reflections all possess a sizable structure factor comparable to the structure factor of the $(031)$ reflection. Thus, the presence or absence of the $(130)$ reflection signals, respectively, a multi- or a single-domain state with the twin orientation favored by the uniaxial pressure. The analysis of the $(\overline{1}\overline{3}0)$ reflection, the Friedel equivalent of $(130)$, yields the same information.
We also analyzed the $(031)$ reflection, which is allowed in space group Pnma, whereas the corresponding reflections of the other 5 twins are either forbidden or possess a considerably smaller structure factor.
Therefore, the emergence of intensity at this reflection is opposed to that at the $(130)$ and $(\overline{1}\overline{3}0)$ reflections. The intensities of the three analyzed reflections did not change upon cooling from 170 to 10~K in the D9 experiment at zero field, proving that their intensity change cannot be associated with a change of the crystal structure, but must arise from a change of the twinning ratios.
Fig.~2 (a,b) display the intensity at the (130) reflection in a sequence of magnetic-field and temperature variations. Starting at 200~K and $\mu_0H=0$\,T, the (130) intensity signals the presence of 15~\% minority twins. After field cooling in $\mu_0H=9$~T to 2~K, this  (130) intensity was fully suppressed, i.e., the sample transformed to the majority twin. On heating in 9~T to 200~K,  the (130) intensity first partially reappears and then fully recovers at 230~K and $\mu_0H=0$\,T.
Figure~\ref{Twins}(c) presents the intensity changes of the $(130)$ reflection together with those of the $(\overline{1}\overline{3}0)$ and $(031)$  reflections as a function of this sequence of 4 points in the magnetic-field and temperature space. The intensities of $(130)$ and of its Friedel equivalent $(\overline{1}\overline{3}0)$ reflection well agree to each other, whereas the intensity of the $(031)$ reflection shows the opposite behavior. Thus, all three reflections consistently reveal the systematic magnetic-field and temperature dependence of the twinning ratio. At low temperature and high magnetic field the structural domains with the $c$ axis parallel to the magnetic field grow on cost of the other domains.
Moreover, there is a shape-memory effect as the field-induced change of the domain fractions is
reversed upon heating to the nonmagnetic phase.

 \begin{figure}
  \includegraphics[width=0.95\columnwidth]{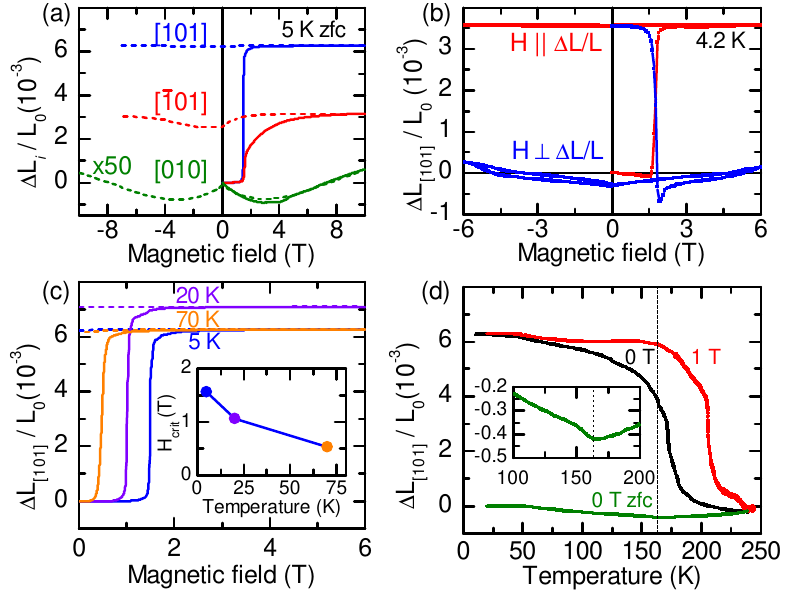}
  \caption{\label{dilatometer} (a) Relative length changes parallel to the magnetic field for the cubic $[110]_c$, $[\overline{1}01]_c$ and the $[010]_c$ direction (the data along $[010]_c$ were multipilied by a factor 50). Solid lines are virgin curves after zero-field cooling (zfc) to 5~K, and dashed lines are obtained during the subsequent field reversals. (b) Switching between the structural domains by applying the external field either along $[101]_{c}$ or along $[\overline{1}01]_{c}$ and measuring $\Delta L(H)/L$ along $[101]_{c}$. (c) $\Delta L(H)/L$ along $[101]_{c}$ for $\vec{H}\| [101]_{c}$ as in (a), but for zfc to $T=5$, 20, or 70~K; the inset shows the reorientation fields of the corresponding virgin curves.  (d) Temperature dependence $\Delta L(T)/L$ measured along $[101]_{c}$ for a multi-domain state obtained by zfc (green, with an enlarged view in the Inset) in comparison to single-domain states, which  were prepared by field poling at 5~K and decreasing the field to 0 (black) or 1~T (red). The vertical line marks $T_c=165$~K of the ferromagnetic ordering.}
  \label{MS}
 \end{figure}

The magnetic-field induced switching of the structural domains also causes strong changes of the macroscopic sample length, which are studied using a capacitance dilatometer. Again, $mm$-size single crystals of almost cubic shapes with edges parallel to the cubic $[101]_{c}$,  $[\overline{1}01]_{c}$, and $[010]_{c}$ directions were used. After zero-field cooling to 5 K, the relative length changes $\Delta L_i(H)/L_0$ were measured along these  3 directions $i$ up to $\mu_0H =9$~T in longitudinal configurations $\vec{H}\| L_i$. We find drastic elongations of the order of several $10^{-3}$ in the virgin curves of $[101]_{c}$ and  $[\overline{1}01]_{c}$, when the field is initially ramped up, whereas much weaker effects are observed when the field is subsequently removed or reversed; see Fig.~\ref{MS}(a). For $\vec{H}\| L_i \| [010]_{c}$,  the behavior is completely different: the overall length change is reduced by about 2 orders of magnitude and the general field dependence is qualitatively different and almost non-hysteretic. These length changes can be naturally explained by magnetic-field induced twin reorientations. The $ [101]_{c}$ and  $[\overline{1}01]_{c}$ directions refer either to the $a$, $c$ or $[121]$ directions of orthorhombic domains. Thus, the magnetic field can pole the originally multi-domain sample to a single-domain state with the magnetic easy axis $c$ aligned along the field. Due to different orthorhombic lattice constants ($c/a = 1.0064$, $\sqrt{2}c/b = 1.0033$ at 10 K\cite{Lee2013}) this poling causes a drastic elongation of several $10^{-3}$ in the respective virgin curves. The subsequent field-sweep cycle only requires a sign reversal of the magnetization  $\vec{M}$, which can be achieved by a switching between $180^\circ$ magnetic domains within a single structural domain. In the simplest view, such a sign reversal could occur without any length change, but one has to consider also the strains arising from the finite width of the domain walls during the magnetization reversal. These effects are small compared to the main effect in the virgin curve. The very weak effect for $\vec{H}\| L_i \| [010]_{c}$ has a similar origin, because this direction does not correspond to the magnetic easy axes of any of the 6 orthorhombic twin domains. The 2 orientations with $b\|\vec{H}$ have their easy axis in the plane perpendicular to the field, while the easy axes of the other 4 twins ($b\perp\vec{H}$) are at 45$^\circ$ with respect to $\vec{H}$. Consequently, the magnetic field cannot induce a structurally single-domain state with the longest axis $c\|\vec{H}$. Instead, the field first mainly reorients the $180^\circ$ magnetic domains within the four $b\perp\vec{H}$ domains and then partially reorients the two $b\|\vec{H}$ domains into $b\perp\vec{H}$.
In addition the magnetization $\vec{M}$ continuously rotates from the respective easy $c$ axis towards $\vec{H}$, as will be further discussed below in the context of the $\vec{M}(\vec{H})$ curves. This partial domain rearrangement, however, results in a much smaller length change, because the average of $a$ and $c$ differs from $b$ by only 10$^{-4}$~\cite{Lee2013}.
The corresponding changes $\Delta L_i(H)/L_0$ thus essentially reflect the intrinsic magnetostriction of SrRuO$_3$ and are in the range of $10^{-5}$ as it is typical for magnetic materials.

Figure~\ref{MS}(b) shows $\Delta L_i(H)$ data, which were obtained with a dilatometer that can be rotated in a split-coil magnet in order to vary the angle between $\vec{H}$ and the measured $\Delta L_i$. For the measurement of $\Delta L_i \| [101]_{c}$ the sample was aligned with $[010]_{c}$ parallel to the rotation axis and we studied the longitudinal ($\vec{H}\|\Delta L_i \| [101]_{c}$) and the transverse ($\vec{H}\|[\overline{1}01]_{c} \perp \Delta L_i\| [101]_{c}$) configuration. Starting with an initially twinned crystal, a single-domain state with either  $c\|[101]_{c}$ (longitudinal) or with $a\|[101]_{c}$ (transverse) is induced and remains stable after the field is removed again. The data of Fig.~\ref{MS}(b) were obtained after such an initial poling and clearly illustrate that the crystal reversibly switches between these 2 differently aligned single-domain states, if the field is subsequently applied and removed either in the longitudinal or the transverse direction. At $T\simeq 4.2$~K, the  switching field amounts to $\mu_0H_s\simeq 1.76$~T in agreement to the corresponding values of Fig.~\ref{MS}(a) obtained on a different crystal. The magnetic-field induced switching  of the structural domains is remarkably sharp and, as shown in Fig.~\ref{MS}(c), the switching field decreases from $\mu_0 H_s\simeq 1.7$~T at 5~K to $\simeq 0.5$~T at 70~K.

Figures~\ref{MS}(c,d) illustrate the magnetic shape-memory effect of the structural domains. In zero field, the twinned crystal exhibits a moderate temperature dependence of $\Delta L_i(T)/L_0$ ($L_i \| [101]_{c}$) with an anomaly at $T_c~\simeq 163$~K, see Inset. After field-poling the crystal at 5~K to a single-domain state, very different $\Delta L_i(T)/L_0$  curves are obtained for $\mu_0H=0$ or 1~T. The temperature dependence of $\Delta L_i(T)/L_0$ is moderate up to about 100~K or 150~K, but then a strong decrease of $\Delta L_i(T)/L_0$ sets in, which becomes most pronounced around 172~K and 205~K for 0 and 1~T, respectively. The single-domain state enforced by the field poling remains thus stable up to rather high temperature, but above the zero-field $T_c$ the initial multi-domain state recovers as in the diffraction experiment, which reveals the magnetic shape-memory effect.

%
%

\begin{figure}
  \includegraphics[width=0.90\columnwidth]{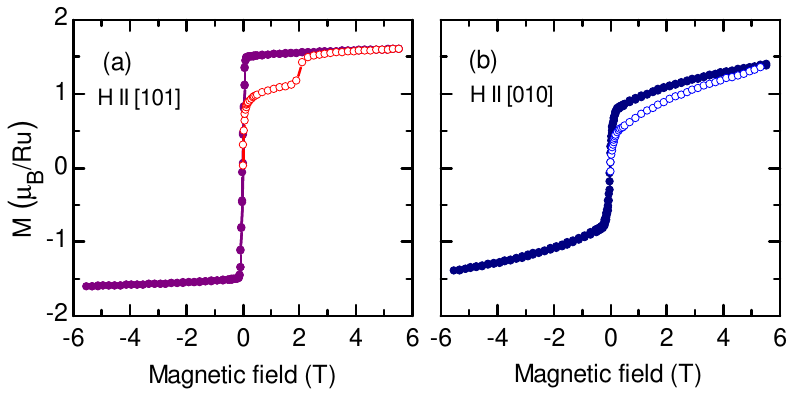}
  \caption{\label{MNeukurve} Magnetization for fields along the cubic $[101]_c$ (a) and $[010]_c$ (b) directions. Virgin curves ($\circ$) are obtained after zero-field cooling to 2~K and followed by full hysteresis cycles ($\bullet$).}
  \label{mag}
 \end{figure}

Figure~\ref{mag} shows the magnetization for $\vec{H}\|[101]_{c}$ and $\vec{H}\|[010]_{c}$ obtained in a commercial SQUID magnetometer (Quantum Design). After zero-field cooling the sample to 2~K, the virgin curve of $M(H)$ for $\vec{H}\|[101]_{c}$ rapidly increases to about $0.9 \mu_{\rm B}$/Ru, then it follows almost a plateau until it shows another step at $\mu_0 H_s\simeq 2\,$T to a saturation value of $M_{\text{sat}}\simeq 1.6 \mu_{\rm B}$/Ru. In the subsequent field-sweep cycle, the intermediate plateau is absent and the $M(H)$ curve essentially switches between the 2 saturation plateaus with almost no hysteresis. This perfectly agrees with the magnetically induced structural change. In the virgin curve, the magnetic moments of the structural domains with $c\|H$ saturate first, then at $\mu_0 H_s\simeq 2\,$T the other structural domains reorient and full saturation is reached. In the subsequent field sweep cycles, the sample remains in the structural single-domain state and $M(H)$ switches between $180^\circ$ magnetic domains.

For $\vec{H}\|[010]_{c}$, the virgin curve of $M(H)$ also rapidly increases, but to a smaller value of $\simeq 0.5 \mu_{\rm B}$/Ru and then $M(H)$ continuously increases to $\simeq 1.4 \mu_{\rm B}$/Ru at $\mu_0 H=6\,$T; see Fig.~\ref{mag}(b). Upon decreasing the field, $M(H)$ does not follow the virgin curve, but approaches a larger remanent magnetization of $M_{\text{rem}}\simeq 0.75 \mu_{\rm B}$/Ru in zero field and finally $M(H)$ follows the reverted branch in the negative field range. As discussed above, this different behavior is related to the fact that $\vec{H}\|[010]_{c}$ does not correspond to the magnetic easy axis of any of the 6 orthorhombic twin orientations. Thus, $M_{\text{rem}}$ stems from the 4 twins, whose easy-axis $c$ is at 45$^\circ$ with respect to  $\vec{H}$ (thus $b\perp\vec{H}$), and the  two $b\|\vec{H}$ orientations with easy-axis perpendicular to $\vec{H}$ do not contribute. With $M_{\text{sat}}\simeq 1.6 \mu_{\rm B}$/Ru, the occupation of the 4 contributing twin orientations in zero field can be estimated by $\sqrt{2}M_{\text{rem}}/M_{\text{sat}}\simeq 2/3$  yielding equal domain population after the field decrease.
The smaller initial increase in the virgin curve yields a population of only 40\,\% \ for the 4 contributing $b\perp\vec{H}$ orientations, while the 2 non-contributing twins with $b\|\vec{H}$ would be populated by about 60\,\%. This means that the initial field increase causes a sizeable reorientation of structural domains, but this reorientation is rather gradual. These gradual changes most probably result from the fact that the magnetization cants from the respective easy-axis direction towards $\vec{H}$. Thus, for high-enough fields the potential energies $-\mu_0 \vec{M}\vec{H}$ of the differently oriented twins do not vary too much and, as a consequence, their difference does not exceed the necessary elastic energy for a domain reorientation.

From an extrapolation of the measured $M(H)$ for $\vec{H}\|[010]_{c}$ one may estimate the intrinsic anisotropy field to $\mu_0 H_{\text{an}}\approx 10$\,T, which agrees with the energy of the ferromagnetic resonance observed in time-resolved magneto-optical Kerr-effect measurements~\cite{Langner2009}. A similar intrinsic anisotropy may be expected for the orthorhombic $a$ direction, but this anisotropy remains hidden by the structural domain reorientation.

The magnetic-field induced control of structural domains in SrRuO$_3$ is remarkable in view of the largely different magnetic and structural transition temperatures. The spin-orbit coupling in this $4d$ material is, however, significantly larger than in the well studied $3d$ oxides. Thus, the elongation of the RuO$_6$ octahedron along $c$ results in a sizeable easy-axis anisotropy with an aniostropy field $\mu_0 H_{\text{an}}\approx 10$\,T. This  prevents magnetic spin-flop transitions in fields of the order of 1\,T, but the magnetic energy $-\mu_0 \vec{H}\vec{M}$ can be gained by structural domain flops. Thus, a strong Ising anisotropy and weak pinning of structural domains is required, which both seem to be fulfilled in SrRuO$_3$. The  involved magnetic energy density can be estimated from $M_{\text{sat}}\simeq 1.6\,\mu_{\rm B}$/Ru, the switching field $\mu_0 H_s\simeq 1.8\,$T of the domain flop, and the volume per formula unit $V_{fu}\simeq 60\,$\AA$^3$ to $E_{mag}=\mu_0 H_sM_{\text{sat}}/V_{fu}\approx 0.4\,$MPa. This small energy density is sufficient to induce a detwinning with macroscopic length changes up to several $ 10^{-3}$ in $\Delta L_i/L_0$, whereas comparable elastic distortions in usual single-domain solids would require energy densities in the GPa range. The field induced domain flop seems to be visible already in very early data of Kanbayasi \cite{Kanbayasi1976a} reporting anomalous switching effects and a tetragonal symmetry in flux-grown single crystals in higher fields. Moreover, the strange angular hysteresis  in the  magnetoresistance of thin-film samples is probably also related to structural domain flops~\cite{Ziese2010}.

In conclusion, the combination of neutron diffraction with macroscopic measurements reveals the magnetic control of structural domains in SrRuO$_3$.
Applying magnetic fields of the order of only 1 T along one of the cubic $[110]_c$ directions, which by symmetry can become an orthorhombic $c$ direction, induces a
structurally single-domain state with the magnetic easy axis $c$ aligned along the field. This single-domain state is stable at low temperature so that successive magnetic-field  cycles yield narrow magnetic hysteresis-loops without further structural domain changes.  Rotating the field to another symmetry-equivalent cubic direction, e.g.\ $[\overline{1}10]_c$, even allows to switch between different symmetry-equivalent single-domain states. Upon heating above the Curie temperature the initial domain arrangement recovers and reveals the magnetic shape-memory effect. Magnetic fields along a cubic $[010]_c$ direction, which cannot become an easy axis, also induce a partial domain reorientation, but the intrinsic
magnetic anisotropy of the order of 10\,T remains visible. The magnetic-field control of structural domains in SrRuO$_3$ arises from the large magnetic anisotropy due to the strong spin-orbit coupling in this $4d$ transition-metal oxide combined with weak pinning of structural domain walls. Similar control of structural domains should also be observable for other ferroic transitions such as ferroelectric instabilities.

We thank I. Lindfors-Vrejoiu for fruitful discussions.
This work was supported by the Institutional Strategy of the University of Cologne within the German Excellence Initiative, CRC 1238 Projects No. A02, No. B01 and No. B04


\begin{thebibliography}{26}
\bibitem [{Pnm()}]{Pnma}%
  \BibitemOpen
  \href@noop {} {}\bibinfo {note} { Throughout the paper we refer to the orthorhombic 
  lattice, unless the cubic system is specified by a subscript.
  Note, that in space group Pnma the
  octahedron rotates around the orthorhombic $b$ axis and tilts roughly around the
  $a$ axis and that $a < c$ indicating an elongation of
  the octahedron basal plane along $c$ in SrRuO$_3$.}\BibitemShut {Stop}%
\bibitem [{D9()}]{D9}%
  \BibitemOpen
  \href@noop {} {}\bibinfo {note} {Two sets of reflections were recorded at 170
  and 10~K consisting of 1147 and 1642 reflections, respectively.
  Friedel-equivalent reflections were averaged yielding 617 and 833
  symmetrically different ones with R$_{int}$ values of 2.28 and 2.33 \% for
  the datasets. A gaussian extinction correction of type I is applied. The
  refinements yield R values of $wR(all)=4.57$ and 4.3~\% for 10 and 170~K,
  respectively.}\BibitemShut {Stop}%
\bibitem [{Str()}]{StructuralParametersSrRuO3}%
  \BibitemOpen
  \href@noop {} {}\bibinfo {note} {The structural parameter $x$,$y$,$z$ and
  atomic displacement parameters $U_{ij}$ in 10$^{-4}$ \AA$^2$ at 10\,K are for 
  Sr:  $0.0220(3)$, $0.25$, $-0.0036(5)$, $56(8)$, $25(9)$, $45(9)$, 0, 3(5), 0 
  Ru: $0$,$0$, $0.5$, $ 60(9)$, $ 44(9)$, $ 15(8)$, 12(5), 25(12), -3(7)
  O1: $-0.0046(3)$, $0.25$, $0.4460(4)$, $42(9)$, $0(8)$, $64(9)$, 0, 10(4), 0
  O2: $0.2779(4)$, $0.02871(14)$, $0.07220(3)$, $103(10)$, $69(7)$, $73(10)$ , 7(6), 3(3), 17(5)
and at 170~K for 
Sr:   $0.0206(3)$, $0.25$, $-0.0031(6)$, $79(8)$, $37(14)$, $93(10)$, 0, 2(5), 0
Ru: $0$,  $0$, $0.5$, $ 66(11)$, $ 117(18)$, $ 26(11)$, 6(5), 34(15), 15(7)
O1: $-0.0032(3)$, $0.25$, $0.4472(5)$, $53(12)$, $-25(14)$, $78(11)$, 0, 4(4), 0
O2: $0.2789(5)$, $0.0299(3)$, $0.7229(6)$, $122(10)$, $105(9)$, $119(10)$, -23(11), 0(4), 8(11)}\BibitemShut
  {NoStop}%
\end{thebibliography}
\end{document}